# Nonlinear valley and spin currents from Fermi pocket anisotropy in 2D crystals

**Authors:** Hongyi Yu[1], Yue Wu[1], Gui-Bin Liu[2,1], Xiaodong Xu[3,4], Wang Yao[1*]

**Affiliations:**

[1]Department of Physics and Center of Theoretical and Computational Physics, University of Hong Kong, Hong Kong, China

[2]School of Physics, Beijing Institute of Technology, Beijing 100081, China

[3]Department of Physics, University of Washington, Seattle, Washington 98195, USA

[4]Department of Materials Science and Engineering, University of Washington, Seattle, Washington 98195, USA

*Correspondence to: wangyao@hku.hk

**Abstract:**

**Controlled flow of spin and valley pseudospin is key to future electronics exploiting these internal degrees of freedom of carriers. Here we discover a universal possibility for generating spin and valley currents by electric bias or temperature gradient only, which arises from the anisotropy of Fermi pockets in crystalline solids. We find spin and valley currents to the second order in the electric field, as well as their thermoelectric counterparts, i.e. the nonlinear spin and valley Seebeck effects. These second-order nonlinear responses allow two unprecedented possibilities to generate pure spin and valley flows without net charge current: (i) by an AC bias; or (ii) by an arbitrary inhomogeneous temperature distribution. As examples, we predict appreciable nonlinear spin and valley currents in two-dimensional (2D) crystals including graphene, monolayer and trilayer transition metal dichalcogenides, and monolayer gallium selenide. Our finding points to a new route towards electrical and thermal generations of spin and valley currents for spintronic and valleytronic applications based on 2D quantum materials.**

The discovery of atomically thin two-dimensional (2D) crystals has opened up new realms in physics, material science, and engineering [1,2]. The library of 2D crystals now consists of versatile members including graphene and its derivatives, oxides, and transition metal dichalcogenides (TMDs), offering a variety of appealing material systems, from gapless to direct-gap semiconductors, and from metal to wide-gap insulators [1-3]. A rather common feature of these 2D crystals is the presence of the conduction and valence band edges at degenerate extrema in momentum space, usually referred to as valleys. The Fermi surface then consists of well-separated pockets at the valleys, which constitute an effective internal degree of freedom of the carrier. The exploitation of the valley pseudospin, as well as spin, in electronics may significantly extend the device functionalities [4-9]. When spin-orbit interaction splits the band, the spin splitting must have opposite sign in a pair of valleys related by time reversal symmetry, realizing an effective coupling between the spin and valley pseudospin [10]. The recent discoveries of valley physics and spin-valley coupled effects in 2D TMDs have significantly boosted their potential in spintronic and valleytronic applications [10-18].

The generation and control of spin and valley pseudospin currents are at the heart of spintronics and valleytronics [19]. There has been a variety of approaches based on the detail characteristics of different systems, for example, the spin injection or pumping from proximity ferromagnets [20,21], and the various optical injection methods that rely on optical selection rules [22-24]. In time reversal symmetric systems, the spin Hall effect from spin-orbit coupling [25-28] and the valley Hall effect from inversion symmetry breaking [6,13] have also been explored, with possibility of implementation in 2D crystals [18,29,30]. The spin or valley Hall current, however, is always accompanied by the longitudinal charge current that is orders of magnitude larger, and such a major cause of dissipation cannot be removed as it has the same linear dependence on the field as the Hall currents.

Here we discover a new origin of valley and spin currents from the anisotropy of Fermi pockets, a universal feature of crystalline solids. Such valley and spin currents can be generated by the electric bias only, and appear in the second order to the electric field. The direction of the valley and spin currents is controlled by the relative orientation of the field to the crystalline axis. The quadratic dependence on field makes possible current rectification for generation of dc spin and valley currents by ac electric field, with the absence of net charge current. For several exemplary 2D crystals including TMDs monolayers and trilayers, graphene, and GaSe monolayer, we find appreciable nonlinear spin and valley currents in their $K$, $\Gamma$, and $\Lambda$ valleys, where the ratio of the valley current to the charge current has a simple dependence on the degree of band anisotropy. In monolayer TMDs, our estimation shows that the nonlinear valley (spin) current starts to exceed the observed sizable valley (spin) Hall current [13,18], at a small electric field of $\sim 10$ mV·μm$^{-1}$. We predict that, at p-n junction in monolayer TMDs [31-33], the nonlinear valley current will result in unique circular polarization pattern of electroluminescence depending on the orientation of the junction relative to the crystalline axis. We also predict the nonlinear valley and spin Seebeck effects from the Fermi pocket anisotropy, where a temperature gradient can play the same role as the electric field in giving rise to the nonlinear valley and spin currents. The quadratic dependence of the valley (spin) thermopower on the temperature gradient implies a remarkably simple way to generate pure valley (spin) flow with zero charge current by an inhomogeneous temperature distribution.

**General theory**. We focus here on 2D crystals with mirror symmetry in the out-of-plane (*z*) direction. With this symmetry, the Bloch states must have their spin either parallel or antiparallel to the *z*-axis. Consider a spin-up Fermi pocket in valley $A$ with dispersion $\mathcal{E}_{A,\uparrow}(\boldsymbol{q})$, $\boldsymbol{q}$ being the wavevector measured from $A$. In an in-plane electric field $\boldsymbol{E}$, $f(\boldsymbol{q},\boldsymbol{E})$ is the steady-state distribution function of carriers, and the current is then: $\boldsymbol{j}_{A,\uparrow}(\boldsymbol{E}) = \int d\boldsymbol{q}\, f_{A,\uparrow}(\boldsymbol{q},\boldsymbol{E})\nabla_{\boldsymbol{q}}\mathcal{E}_{A,\uparrow}(\boldsymbol{q})$. We do *not* concern any Hall current here. For anisotropic dispersion, the conductivity tensor can depend on the field direction. In particular, if $\mathcal{E}_{A,\uparrow}(\boldsymbol{q}) \neq \mathcal{E}_{A,\uparrow}(-\boldsymbol{q})$, the current response can also lack the 180° rotational symmetry, i.e. $\boldsymbol{j}_{A,\uparrow}(\boldsymbol{E}) \neq -\boldsymbol{j}_{A,\uparrow}(-\boldsymbol{E})$. In presence of time-reversal symmetry, this current will have a counterpart $\boldsymbol{j}_{\bar{A},\downarrow}$ from a spin-down pocket at valley $\bar{A}$, the time reversal of $A$, where $\mathcal{E}_{\bar{A},\downarrow}(\boldsymbol{q}) = \mathcal{E}_{A,\uparrow}(-\boldsymbol{q})$ and consequently $f_{\bar{A},\downarrow}(\boldsymbol{q},\boldsymbol{E}) = f_{A,\uparrow}(-\boldsymbol{q},-\boldsymbol{E})$. These determine $\boldsymbol{j}_{\bar{A},\downarrow}(\boldsymbol{E}) = -\boldsymbol{j}_{A,\uparrow}(-\boldsymbol{E})$ (c.f. Fig. 1). Thus, under the condition of Fermi pocket anisotropy, the currents contributed by the time reversal pair of Fermi pockets can have a finite difference: $\boldsymbol{j}_{A,\uparrow}(\boldsymbol{E}) - \boldsymbol{j}_{\bar{A},\downarrow}(\boldsymbol{E}) \neq 0$, which is a valley current as well as a spin current.

We find that such spin and valley currents arise in the second order of the electric field. In an electric field along the *x*-direction, without concerning the Hall effect, the longitudinal and transverse components of $\boldsymbol{j}_{A,\uparrow}$ can be expanded as [34]:

$$j^x_{A,\uparrow}(\boldsymbol{E}) = \sigma^{xx}_{A,\uparrow}E + \sigma^{xxx}_{A,\uparrow}E^2 + O(E^3), \qquad j^y_{A,\uparrow}(\boldsymbol{E}) = \sigma^{yxx}_{A,\uparrow}E^2 + O(E^3). \tag{1}$$

As $j_{\bar{A},\downarrow}(-E) = -j_{A,\uparrow}(E)$, we have $\sigma_{A,\uparrow}^{xxx} = -\sigma_{\bar{A},\downarrow}^{xxx}$, $\sigma_{A,\uparrow}^{yxx} = -\sigma_{\bar{A},\downarrow}^{yxx}$, while $\sigma_{A,\uparrow}^{xx} = \sigma_{\bar{A},\downarrow}^{xx}$. The charge current is $j_{A,\uparrow}(E) + j_{\bar{A},\downarrow}(E) = 2\hat{x}\sigma_{A,\uparrow}^{xx}E + O(E^3)$, an odd function of the electric field, while the valley (spin) current is $j_{A,\uparrow}(E) - j_{\bar{A},\downarrow}(E) = 2(\hat{x}\sigma_{A,\uparrow}^{xxx} + \hat{y}\sigma_{A,\uparrow}^{yxx})E^2$, an even function of the field.

The distinct dependence of charge and valley (spin) currents on the electric field leads to the *rectification* of spin and valley currents with vanishing net charge current. Applying an ac electric field $E_x = E \cos \omega t$, the dc charge current is zero, and the valley (spin) current becomes

$$j_{A,\uparrow} - j_{\bar{A},\downarrow} = (\hat{x}\sigma_{A,\uparrow}^{xxx} + \hat{y}\sigma_{A,\uparrow}^{yxx})E^2(1 + \cos 2\omega t). \tag{2}$$

In addition to a second harmonic term, the valley (spin) current has a dc component. We note that Eq. (2) implicitly assumes $\omega^{-1}$ being larger than the momentum relaxation time $\tau$, as it is based on the steady state response in Eq. (1). From the symmetry alone, we expect this rectification effect can exist even beyond the regime of $\omega\tau < 1$.

**Nonlinear valley and spin currents in monolayer TMDs**. Monolayer (ML) group-VIB TMDs provide an excellent system to illustrate the different scenarios of the nonlinear spin and valley currents (c.f. Fig. 2). In such hexagonal 2D crystals, the center ($\Gamma$) and the corners ($K$ and $\bar{K}$) of the hexagonal Brillouin zone are high symmetry points where band extrema are expected. The top valence band in ML TMDs has local maxima at both $\Gamma$ and $K$ ($\bar{K}$) points. The lowest conduction band has two types of local minima: the $K$ ($\bar{K}$) point, and the low-symmetry $\Lambda$ ($\bar{\Lambda}$) points between $K$ ($\bar{K}$) and $\Gamma$.

For the Fermi pockets at $K$ ($\bar{K}$), the anisotropy is the trigonal warping [35,36], which breaks the $180°$ rotational symmetry of the pockets. Both the conduction and the valence bands are spin split in the $K$ valleys [13,37]. The spitting is $O(100)$ meV in the valence band, and $O(10)$ meV in the conduction band. If the Fermi energy is between the split bands, we only have a spin up (down) Fermi pocket at $K$ ($\bar{K}$). Valley current is then the same as spin current. If the field is applied along a zigzag direction, the valley (spin) current is either parallel or antiparallel to the field because of the reflection symmetry of Fermi pocket (c.f. Fig. 2a). For electric field in armchair direction, we find the valley (spin) current perpendicular to the field (see Fig. 2d-e).

The trigonal warping also exists for the hole pocket at $\Gamma$ point. By the time reversal symmetry, the warping is opposite for the spin up and down pockets (c.f. Fig. 2b), giving rise to a nonlinear spin current. At the six low-symmetry Λ valleys in the conduction band (Fig. 2c), the anisotropy leads to valley-dependent current response to electric field, as well as an overall spin current contributed by all Λ and $\bar{\Lambda}$ pockets. The direction of spin current from Γ or Λ pockets as a function of field orientation is also similar to the $K$ pockets (c.f. Fig. 2e and Table 1).

**Quantitative results for TMDs, GaSe and graphene**. The magnitude of the nonlinear spin and valley currents depends on the dispersion of the Fermi pockets and the distribution function in electric field. For the latter, we adopt the commonly used relaxation time approximation. Consider for example the $K$ valleys in ML TMDs, the dispersion of either the electron or the hole near the Fermi surface can be well fit by [35,36]:

$$\mathcal{E}_K(\boldsymbol{q}) = \frac{\hbar^2 q^2}{2m^*}(1 + \beta q \cos 3\theta_q), \qquad (3)$$

where $\boldsymbol{q} \equiv (q \cos\theta_q, q \sin\theta_q)$, $\beta$ has weak dependence on the Fermi energy [34]. Neglecting the spin and valley relaxations, the spin and valley currents from the $K$ and $\bar{K}$ pockets are (c.f. supplementary material [34]):

$$\boldsymbol{j}_s = \boldsymbol{j}_v = \frac{12\pi}{\hbar}\mathcal{E}_F \beta |\boldsymbol{k}_d|^2 (\cos 2\theta, -\sin 2\theta) + O(|\boldsymbol{k}_d|^4), \qquad (4)$$

where $\mathcal{E}_F$ is the Fermi energy measured from the band edge, $\boldsymbol{k}_d = e\tau \boldsymbol{E}/\hbar$ is the displacement of the Fermi surface by the electric field, $\boldsymbol{E} \equiv (E\cos\theta, E\sin\theta)$ and $\tau$ is the momentum relaxation time. $\beta$ measures the degree of anisotropy on the Fermi surface. Comparing this with the reported valley (spin) Hall current in monolayer TMDs [13,18], we estimate that the nonlinear valley (spin) current starts to dominate at a small electric field of $\sim 10 \text{ mV} \cdot \mu\text{m}^{-1}$.

The charge current normalized by $e$ is $\boldsymbol{j}_c = \frac{4\pi}{\hbar}\mathcal{E}_F \boldsymbol{k}_d + O(|\boldsymbol{k}_d|^3)$. The ratio of the spin and valley currents to the charge current is

$$j_s/j_c = j_v/j_c = 3\beta\, e\tau E/\hbar. \qquad (5)$$

Interestingly, this ratio is independent of $\mathcal{E}_F$. We note that Eq. (4) is for the situation where $\mathcal{E}_F$ lies between the spin split bands (c.f. Fig. 2a). This is always the case for p-doped ML TMDs because of the giant spin splitting. For n doping, if the higher spin split band is also occupied, it will have a contribution also given by Eq. (4), but with $\boldsymbol{j}_s = -\boldsymbol{j}_v$ [34]. Eq. (4) still holds for the valley current, but the overall spin current can then differ from the valley current, as listed in Table 1.

Similar analysis can be performed both for Fermi pockets at other band extrema in ML TMDs and for other 2D crystals. For the Γ hole pockets in ML TMDs, the dispersion can be described by Eq. (3) as well, which leads to the spin current given by Eq. (4). The Λ electron pockets in ML TMDs have more complicated dispersion. Nevertheless, the overall spin current from all Λ and $\bar{\Lambda}$ pockets is still given by Eq. (4) (c.f. supplementary material [34]). The values of the parameter $\beta$ for the $K$, Γ, and Λ pockets obtained by fitting the *ab initio* bands are listed respectively in Table 1 for ML $MoS_2$. The corresponding values of $\beta$ in other three ML TMDs, i.e. $MoSe_2$, $WS_2$, $WSe_2$, are found to have comparable magnitudes (see Table S1 in [34]). In ML TMDs, the Γ and Λ pockets only appear at very large p- and n- doping respectively. In trilayer TMDs, which have the same symmetries as the monolayer, Γ and Λ can be the valence and conduction band edges respectively, and we find nonlinear spin and valley currents given by Eq. (4) as well [34]. Table 1 also listed the nonlinear spin current in p-doped monolayer GaSe, where the Fermi pockets are at the Λ points [38], and the result is similar to the TMDs.

Graphene is an example with two representative differences from the scenarios discussed above. First, the bands are spin-degenerate so that spin current must vanish. Second, the band dispersion is linear to the leading order. The conduction and valence bands dispersion at the $K$ and $\bar{K}$ valleys are described by: $\mathcal{E}_K(\boldsymbol{q}) = \pm\hbar v_F q \left(1 + \beta q \cos 3\theta_q + O(q^2)\right)$. Such dispersion can lead to valley dependent tunneling at potential barriers [39,40]. Interestingly, we find that, in graphene, the nonlinear valley current is still given by Eq. (4), and the ratio of the valley current to charge current given by Eq. (5) [34].

Eq. (4) and (5) is derived for the low temperature regime where the Fermi energy $\mathcal{E}_F \gg k_B T$. Beyond this regime, the nonlinear spin and valley currents will depend on temperature. Nevertheless, Eq. (5) for the valley to charge current ratio will still hold, as this ratio is nearly independent of $\mathcal{E}_F$ and hence the filling of the states in equilibrium [34]. The temperature dependence of this ratio comes in through the momentum relaxation time $\tau$.

**Polarized EL from ML TMD p-n diodes**. The emerging monolayer and multilayer TMD p-n junction devices [31-33] provide an ideal laboratory for the exploration of nonlinear valley and spin currents. Under forward bias, electrons (holes) from the $K$ valleys in the n (p) region will reach the junction and produce electroluminescence (EL) through recombination. If the junction is along the armchair direction, the nonlinear valley (spin) current is collinear to the charge current, and carriers accumulated in the junction region are valley polarized. With the valley dependent optical selection rule [7,13-17], we expect the EL will have an overall circular polarization (Fig. 3a). The magnitude of the EL polarization is given by the ratio between valley and charge currents. Given a reasonable forward bias $E \sim 10 \text{ V} \cdot \mu m^{-1}$ at p-n junction, the EL polarization is estimated to be ~ 20% by extrapolating the result of Eq. (5). The EL polarization changes sign when the p-n junction flips (Fig. 3a). This nonlocal valley transport effect is in qualitative agreement with the polarized EL reported very recently in thin flake $WSe_2$ p-n junctions [41]. A quantitative estimation of the EL polarization in the above devices calls for formulations in large field limit, since Eq. (5) is derived for small displacement of the Fermi surface.

Our theory also predicts a unique spatial pattern of EL polarization when the junction is not along the armchair direction, which distinguishes it from other possible mechanisms for the polarized EL [13,18,41]. Consider a p-n junction along the zigzag direction, the nonlinear valley (spin) current is perpendicular to the charge current, and carriers will accumulate with opposite valley polarizations at the two sides. The EL on the two sides of the junction will then have opposite circular polarization. This spatial dependence clearly distinguishes the nonlocal valley transport here from the effect of local change in the population of recombining electrons and holes by the electric field at depletion region proposed in Ref [41]. The effect of the nonlinear valley current is also distinct from the valley Hall current in ML TMDs [6,13,18]. When p-n junction flips sign, the EL polarization on the two sides will change sign if it arises from the valley Hall effect, but will remain unchanged if it is from the nonlinear valley current (Fig. 3b).

**Nonlinear spin and valley Seebeck effects**. Finally, we discuss the currents driven by a temperature gradient $\nabla T$. Similar to that in an electric field, we find the second-order nonlinear response to the temperature gradient is a pure valley (spin) current arising from the Fermi pocket anisotropy, while the linear response is a charge current (c.f. supplementary material [34]). Taking the $K$ pockets in ML TMDs for example, the direction of nonlinear valley current is also given by Fig. 2e, where $\theta$ now represents the relative angle between the direction of $\nabla T$ (green arrows) and a zigzag axis. If $\frac{T}{|\nabla T|}$ is much larger than the mean free path, we find the ratio between the valley and charge currents [34]:

$$j_v/j_c = \frac{6}{\hbar}\alpha\beta k_B|\nabla T|\tau, \tag{6}$$

where the dimensionless coefficient $\alpha$ is a function of $\frac{\varepsilon_F}{k_B T}$ only, as shown in Fig. 4a.

In the low temperature regime $\varepsilon_F \gg k_B T$, we find $\alpha \cong \frac{\varepsilon_F}{k_B T}$ (c.f. Fig. 4a), and the nonlinear valley current is given by:

$$\boldsymbol{j}_v(\nabla T) = \frac{8\pi^3}{h^3}\varepsilon_F \beta k_B^2 |\nabla T|^2 \tau^2 (\cos 2\theta, -\sin 2\theta) \tag{7}$$

Interestingly, comparing this with Eq. (4), we find

$$\left(\frac{1}{k_B|\nabla T|}\right)^2 \boldsymbol{j}_v(\nabla T) = \frac{2\pi^2}{3}\left(\frac{1}{eE}\right)^2 \boldsymbol{j}_v(\mathbf{E}) \tag{8}$$

This simple relation between the second order response to the electric field and that to the temperature gradient holds true for the other cases of nonlinear spin and valley currents discussed in Table 1.

The quadratic dependence of valley and spin currents on the temperature gradient makes possible the generation of valley and spin flows in the absence of charge current. Consider an arbitrary temperature distribution, where the temperatures at the two ends of the device equal so that there is no charge current. The valley (spin) current is finite as long as the temperature distribution is inhomogeneous, as the valley (spin) thermopower by the positive and negative temperature gradients have the same sign (see Fig. 4b). This is an unprecedentedly simple way for generating pure valley and spin flows.

**Acknowledgments:** This work is mainly supported by the Croucher Foundation under the Croucher Innovation Award, the Research Grant Council (HKU705513P, HKU9/CRF/13G) and the University Grant Committee (AoE/P-04/08) of Hong Kong SAR. G.B.L. is also supported by the NSFC (11304014), the 973 Program (2013CB934500) of China, and the BIT Basic Research

Funds (20131842001, 20121842003). X.X. is supported by US DoE, BES, Materials Sciences and Engineering Division (DE-SC0008145), NSF (DMR-1150719), and Cottrell Scholar Award.

**Tables**

|  | monolayer MoS$_2$ | | | | trilayer MoS$_2$ | | GaSe | graphene |
|---|---|---|---|---|---|---|---|---|
|  | $K$, h | $K$, e * | $\Lambda$, e | $\Gamma$, h | $\Lambda$, e * | $\Gamma$, h | $\Lambda$, h * | $K$, e (h) |
| $j_s \left(\frac{12\pi}{\hbar}\right)$ | $\beta\mathcal{E}_F k_d^2$ | $\beta\Delta k_d^2$ | $3\beta\mathcal{E}_F k_d^2$ | $\beta\mathcal{E}_F k_d^2$ | $3\beta\Delta k_d^2$ | $\beta\mathcal{E}_F k_d^2$ | $3\beta\Delta k_d^2$ | 0 |
| $j_v \left(\frac{12\pi}{\hbar}\right)$ | $\beta\mathcal{E}_F k_d^2$ | $\beta(2\mathcal{E}_F - \Delta) k_d^2$ | # | n/a | # | n/a | # | $\beta\mathcal{E}_F k_d^2$ |
| $\beta$ (Å) | -0.94 | -0.49 | 0.33 | -0.12 | 0.09 | -0.01 | -1.62 | -0.36 |
| $\theta_{v(s)}$ | $\pi - 2\theta$ | $\pi - 2\theta$ | $-2\theta$ | $\pi - 2\theta$ | $-2\theta$ | $\pi - 2\theta$ | $\pi - 2\theta$ | $\pi - 2\theta$ |

**Table 1** | Spin current ($j_s$) and valley current ($j_v$) in several hexagonal 2D crystals. $k_d \equiv e\tau E/\hbar$, where $\tau$ is the momentum relaxation time. The direction angles of the current ($\theta_{s/v}$) and the field ($\theta$) are both defined with respect to a zigzag axis of the hexagonal crystal.

* For these cases we assume $\mathcal{E}_F$ is larger than the small spin splitting $\Delta$, so that both spin bands are occupied at each valley.

# For $\Lambda$ pockets, the valley current is finite but does not have a unique definition since the degeneracy is larger than 2.

**Figures**

Figure 1

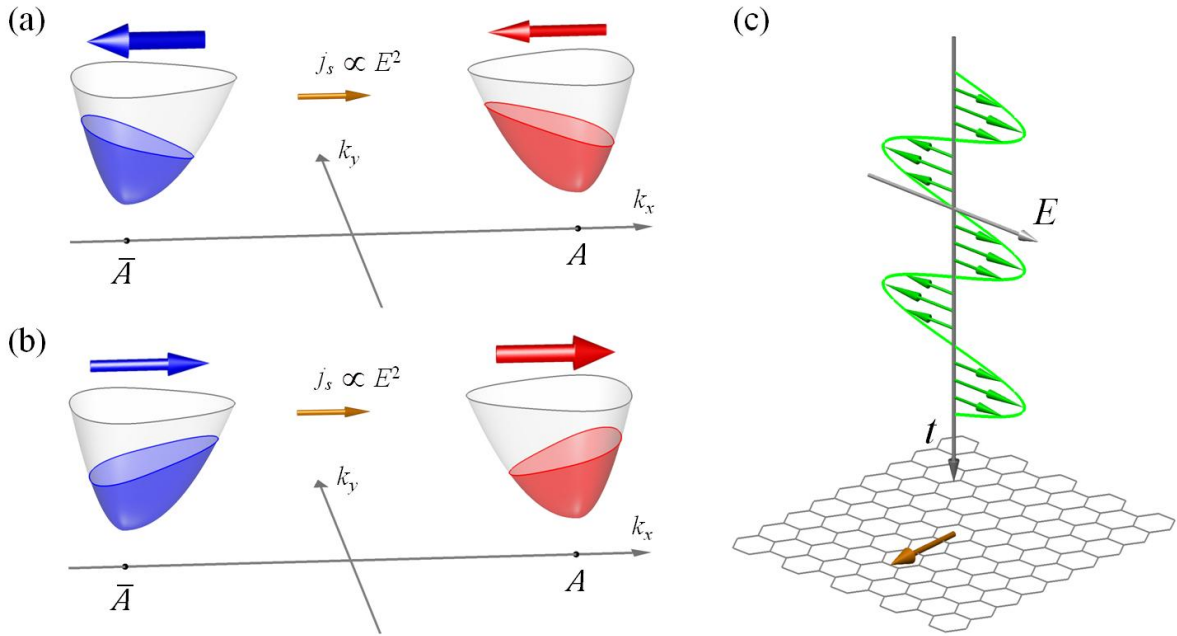

**Figure 1** | (a), (b) Carrier distributions of a spin up Fermi pocket at valley $A$ and a spin down pocket at valley $\bar{A}$, in an electric field along $+x$ (a) or $-x$ (b) direction. The anisotropy of the Fermi pocket results in a difference in the currents from $A$ and $\bar{A}$, giving rise to a valley (spin) current quadratic in the field. (c) Such quadratic dependence in the field makes possible generation of dc valley and spin currents by ac electric field, in the absence of net charge current.

Figure 2

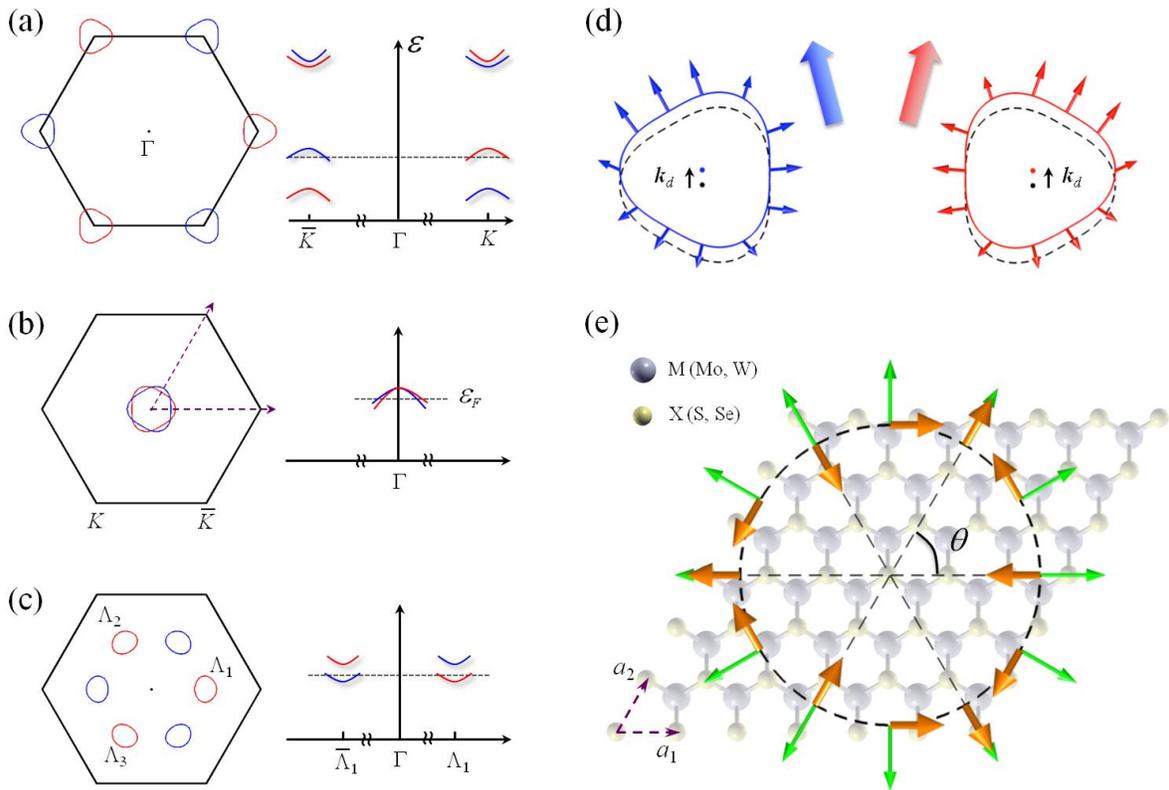

**Figure 2 |** Fermi pockets in 2D TMDs and non-collinear spin and valley currents. (a) Hole pockets at $K$ valleys. (b) Hole pockets at $\Gamma$. (c) Electron pockets at $\Lambda$ valleys. Red and blue denote spin up and down carriers respectively. The dashed horizontal line indicates Fermi level. (d) Displacement of $K$ pockets by an electric field in the armchair direction, where a valley (spin) current flows perpendicular to the field. (e) Dependence of the spin valley current direction (orange arrow) on the relative angle $\theta$ between the field (green arrow) and the crystalline axis.

Figure 3

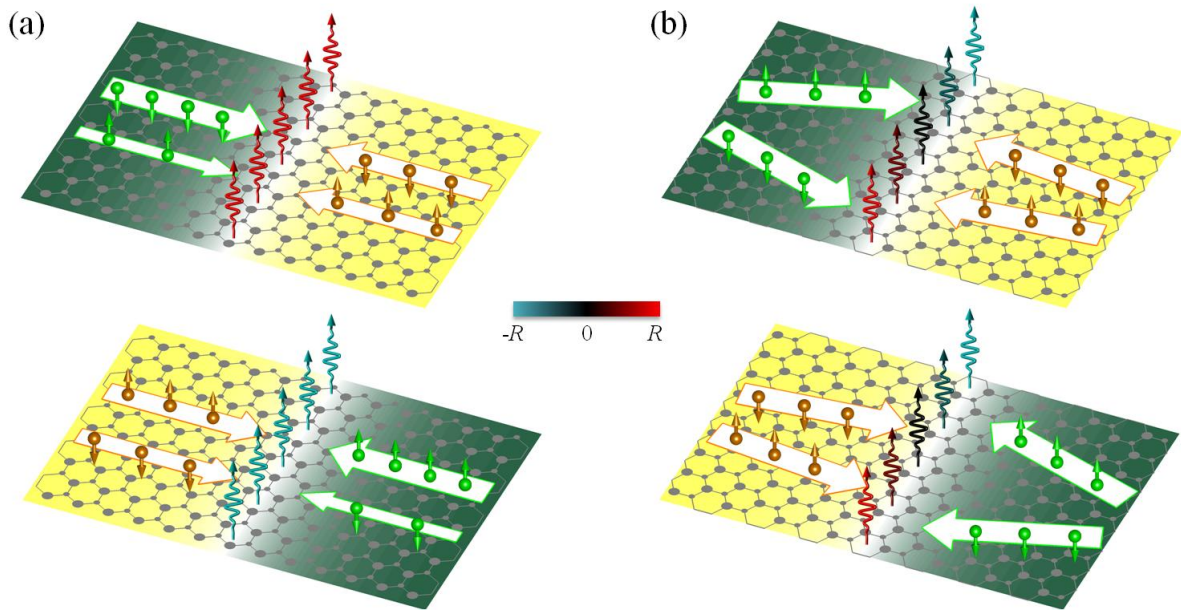

**Figure 3** | (a) Polarized electroluminescence (EL) from p-n junction along armchair direction. The EL has an overall circular polarization $R \sim j_v/j_c$. Green (yellow) color denotes the hole (electron) doped region, and red (blue) color denotes the right- (left-) handed circular polarization. Hole has larger anisotropy, and hence larger nonlinear valley current which determines the EL polarization. The polarization flips sign when the p-n junction flips (top vs. bottom). (b) Spatial pattern of EL polarization from p-n junction along zigzag direction.

Figure 4

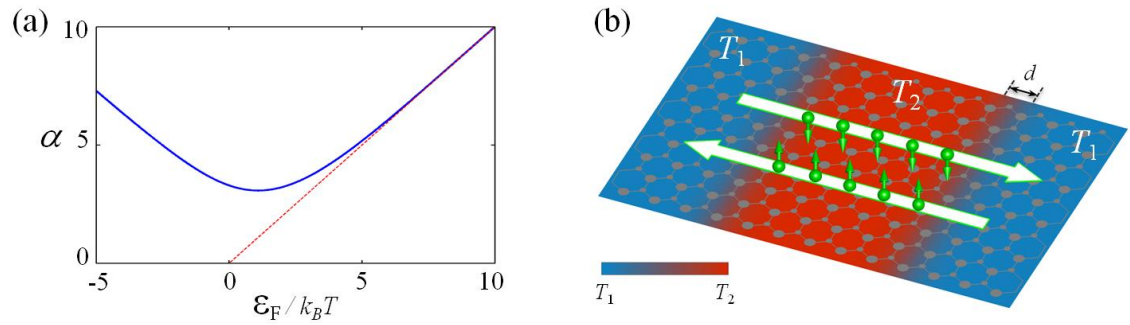

**Figure 4** | (a) The dimensionless coefficient $\alpha$ that measures the ratio between the valley (spin) current and the charge current by a temperature gradient (see Eq. (6)). (b) Spin and valley currents can be generated by an arbitrary inhomogeneous temperature distribution. The charge current vanishes as long as the temperatures at the two ends of the device equal.